\documentclass[prb, twocolumn]{revtex4}
\usepackage{graphicx}
\usepackage{latexsym}
\usepackage{amsmath}
\usepackage{amssymb}
\usepackage{epstopdf}
\usepackage{enumerate}
\usepackage{setspace}   % controllabel line spacing
\usepackage{dcolumn}% Align table columns on decimal point
\usepackage{bm}% bold math
\usepackage{setspace}   % controllabel line spacing

\newcommand{\beq}{\begin{equation}}
\newcommand{\eeq}{\end{equation}}
\newcommand{\beqn}{\begin{eqnarray}}
\newcommand{\eeqn}{\end{eqnarray}}

\begin{document}
\title{Quantum Spin Hall, triplet Superconductor, and topological liquids \\ on the honeycomb lattice}

\author{Cenke Xu}
\affiliation{Department of Physics, University of California,
Santa Barbara, CA 93106}

\begin{abstract}

We classify the order parameters on the honeycomb lattice using
the $\mathrm{SO(4)}$ symmetry of the Hubbard model. We will focus
on the topologically nontrivial quantum spin Hall order and spin
triplet superconductor, which together belong to the $(\bf{3},
\bf{3})$ representation of the SO(4). Depending on the microscopic
parameters, this $(\bf{3}, \bf{3})$ order parameter has two types
of ground states with different symmetries: type $A$, the ground
state manifold is $[S^2\otimes S^2]/Z_2$; and type $B$, with
ground state manifold $\mathrm{SO(3)} \otimes Z_2$. We demonstrate
that phase $A$ is adjacent to a $Z_2\otimes Z_2$ topological phase
with mutual semion statistics between spin and charge excitations,
while phase $B$ is adjacent to a nonabelian phase described by
SU(2) Chern-Simons theory. Connections of our study to the recent
quantum Monte Carlo simulation on the Hubbard model on the
honeycomb lattice will also be discussed.

\end{abstract}

\date{\today}

\maketitle

\section{Introduction and symmetry}

We consider a class of (extended) Hubbard models on a bipartite
lattice at half filling, with the following form: \beqn H &=&
\sum_{<i,j>, \sigma} - t c^\dagger_{i,\sigma}c_{j,\sigma} + H.c. +
U n_{i,\uparrow}n_{i, \downarrow} + H^\prime \cr\cr\cr H^\prime &
= & \sum_{i \in sA, \ j \in sB} t_{ij} c^\dagger_{i,\sigma}c_{j,
\sigma} + H.c. \cr\cr &+& \sum_{i,j} J_{ij}\vec{S}_i\cdot
\vec{S}_j + V_{ij} \vec{T}_i\cdot \vec{T}_j + \cdots\cdots \cr\cr
\vec{T}_i &=& \left( (-1)^i \mathrm{Re}[\Delta]_i, \
(-1)^i\mathrm{Im}[\Delta]_i, \ n_i - 1 \right), \label{hubbard}
\eeqn $\vec{S}_i = \frac{1}{2} c^\dagger_i \vec{\sigma} c_j$ is
the spin operator, $\Delta_i = c^t_i i \sigma^y c_i$ is the
on-site spin singlet Cooper pair. This extended Hubbard model has
a manifest SU(2) spin symmetry. However, if electrons only hop
between two different sublattices (sublattice $A$ and $B$, denoted
as $sA$ and $sB$), after a sublattice depended particle-hole
transformation for spin down electrons: \beqn c_{i,\downarrow}
\rightarrow (-1)^i c^\dagger_{i,\downarrow}, \eeqn the Hamiltonian
is almost unchanged except that $U$ changes sign, and $J_{ij}$
switches with $V_{ij}$. This implies that in addition to the
apparent $\mathrm{SU(2)_{spin}}$ symmetry, this model also has an
$\mathrm{SU(2)_{charge}}$ symmetry that mixes $(c_{i, \uparrow},
(-1)^i c^\dagger_{i, \downarrow})$. Therefore the full symmetry of
this extended Hubbard model is \cite{yangzhang,zhang1991} \beqn
\mathrm{SO(4)} \sim [\mathrm{SU(2)_{spin}} \otimes
\mathrm{SU(2)_{charge}}]/Z_2, \label{so4}\eeqn for arbitrary
parameters in Eq.~\ref{hubbard}. For instance, this SO(4) symmetry
holds for the simplest Hubbard model with only on-site Hubbard
interaction and nearest neighbor electron hopping.

Since the SO(4) symmetry is the full symmetry of the extended
Hubbard model Eq.~\ref{hubbard} on any bipartite lattice, all the
order parameters should be classified in terms of the
representations of the SO(4) Lie algebra. In this work we will
take the honeycomb lattice as an example. Using the notation
introduced in Ref.~\cite{xs2010}, we expand the electron at two
Dirac valleys by $d_{1,2} = e^{i \vec{Q}_{1,2} \cdot \vec{r}} c$
(where $\vec{Q}_{1,2} = \pm (\frac{4\pi}{3\sqrt{3}}, 0) $ are the
wavevectors of the valleys), and introduce Pauli matrices
$\tau^{\alpha}$ and $\mu^{\alpha}$ which act on the sublattice and
valley spaces respectively. Then, after introducing real Majorana
fermions $\zeta_a$ as the real and imaginary parts of
$e^{i\frac{\pi}{4}\tau^x}e^{i\frac{\pi}{4}\mu^x} (d_1, i\tau^y
d_2)^t$, we obtain the continuum Lagrangian for the semimetal
phase \beqn \mathcal{L}_0 = \sum_{a = 1}^8 \bar{\zeta}_a
\gamma_\mu\partial_\mu \zeta_a .\eeqn Here $\mu$ is a 2+1
dimensional spacetime index, and the Dirac $\gamma$ matrices are
$(\gamma_0, \gamma_1, \gamma_2) = \tau^y,\tau^z, \tau^x$,
$\bar{\zeta} = \zeta^t \gamma^0$. Using this notation, in the low
energy field theory, the $\mathrm{SU(2)_{spin}}$ and
$\mathrm{SU(2)_{charge}}$ symmetries are generated by the
following matrices \cite{xs2010}: \beqn && S^x = \sigma^x
\rho^y~~,~~S^y = \sigma^y~~,~~S^z = \sigma^z \rho^y, \cr\cr && T^x
= \sigma^y \rho^z~~,~~T^y = \sigma^y \rho^x ~~,~~T^z = \rho^y.
\label{algebra}\eeqn $\sigma^a$ are spin Pauli matrices, while
$\rho^a$ are Pauli matrices that mix the real and imaginary parts
of electron. Notice that $\mathrm{SU(2)\otimes SU(2)}$ is a double
covering of SO(4), which leads to the $Z_2$ in Eq.~\ref{so4}.
%However, this $Z_2$ structure does not show up in the Lie Algebra
%of Eq.~\ref{algebra}.

Based on the symmetry Eq.~\ref{so4} and Lie Algebra
Eq.~\ref{algebra}, the spin and charge are dual to each other for
a large class of extended Hubbard model. This spin-charge duality
will lead to many interesting results in our paper. The structure
of this paper is as following: in section \ref{secGL}, we show
that the two types of topological orders, the quantum spin Hall
(QSH) order and triplet superconductor (T-SC) are unified as one
representation of the SO(4) group, and the Ginzburg-Landau theory
gives two types of ground states with different symmetry
breakings. Section III and IV study the phase diagrams driven by
proliferating the topological defects in the two types of orders
described in section II respectively. In both section III and IV,
we will first give an argument of the phase diagram based on the
quantum numbers of the topological defects, then a more solid
description based on the Majorana liquid formalism developed in
Ref.~\cite{xs2010} will be presented, and the results from these
two approaches match perfectly with each other. Section V
discusses the situation with $\mathrm{SU(2)_{charge}}$ broken down
to $\mathrm{U(1)_{charge}}$ symmetry.

\section{$\mathrm{SO(4)}$ classification and Ginzburg-Landau formalism}
\label{secGL}

Using the SO(4) algebra in Eq.~\ref{algebra}, we will classify the
order parameters which immediately open up a mass gap for the
Dirac fermion in the semi-metal phase. Some simple Dirac mass gap
order parameters can be classified with these symmetries
straightforwardly. For instance, the quantum Hall order parameter
$\bar{\zeta}\zeta$ is a $(\bf{1}, \bf{1})$ representation of
$\mathrm{SO(4)}$ $i.e.$ it is a singlet of both SU(2) symmetries.
The two sublattice N\'{e}el order $N^a = \bar{\zeta} S^a \mu^y
\zeta$ is a $(\bf{3}, \bf{1})$ representation. The fermion
bilinear $M^a = \bar{\zeta} T^a \mu^y \zeta $ which belongs to the
$(\bf{1}, \bf{3})$ representation is a two sublattice
charge-density wave (CDW) and $s-$wave superconductor: $M^z \sim
(-1)^i (n_i - 1)$, $M^x + i M^y \sim c^t_i i\sigma^y c_i$. $M^a$
can be viewed as the spin-charge dual version of $N^a$. In the
simplest Hubbard model, $N^a$ and $M^a$ orders can be realized in
the two limits $U \gg |t|$ and $- U \gg |t|$ respectively.

\begin{figure}
\includegraphics[width=2.0 in]{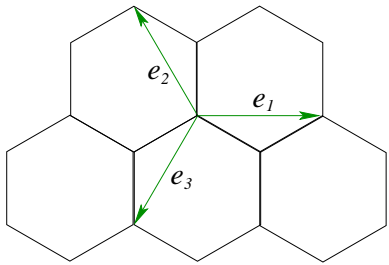}
\caption{Honeycomb lattice and the vectors $e_a$.}
\label{honeycomb}
\end{figure}

Now we discuss the following order parameters which belong to the
$(\bf{3}, \bf{3})$ representation of SO(4): \beqn && Q_{ab} =
\bar{\zeta} A_{ab} \zeta, \cr\cr && A_{ab} = T^aS^b = \left(
\begin{array}{cccc}
- \sigma^z\rho^x & \rho^z & \sigma^x\rho^x \\ \\
\sigma^z\rho^z & \rho^x & - \sigma^x \rho^z \\ \\ \sigma^x &
\sigma^y\rho^y & \sigma^z
\end{array}
\right) \label{Q}\eeqn This $3 \times 3$ matrix $Q_{ab}$ has
$\mathrm{SO(3)_{left}}$ and $\mathrm{SO(3)_{right}}$
transformations, which correspond to $\mathrm{SU(2)_{charge}}$ and
$\mathrm{SU(2)_{spin}}$ symmetry respectively. $Q_{3b}$
corresponds to the QSH vector \cite{kane2005a,kane2005b}, while
$Q_{2b} + i Q_{1b}$ is the spin triplet pairing between next
nearest neighbor sites: \beqn  Q_{3b} &\sim& \sum_{j \in sA, \ a =
1, 2, 3} i c^\dagger_{j} \sigma^b c_{j + e_a} + H.c. \cr\cr &-&
(sA \rightarrow sB) \cr\cr Q_{2b} + i Q_{1b} &\sim& \sum_{j \in
sA, \ a = 1, 2, 3} i c^t_{j} i\sigma^y \sigma^b c_{j + e_a} \cr\cr
&+& (sA \rightarrow sB). \eeqn $e_1 = \sqrt{3}\hat{x}$, $e_2, \
e_3 = - \frac{\sqrt{3}}{2}\hat{x} \pm \frac{3}{2}\hat{y}$ are
three vectors on the honeycomb lattice that connect next nearest
neighbor sites. %The triplet pairing $Q_{2b} + i Q_{1b}$ has a
%similar structure as phase B of Helium$^3$.
Therefore the two types of topological order parameters, QSH and
T-SC, are unified through the $\mathrm{SO(4)}$ symmetry. Under
time reversal symmetry $\mathcal{T}$, $Q_{3b}$ and $Q_{2b}$ are
even, while $Q_{1b}$ is odd. Under reflection symmetry $P_x: y
\rightarrow - y$, $Q_{1b}$ and $Q_{2b}$ are even, while $Q_{3b}$
is odd; under $P_y: x \rightarrow -x$, all components of $Q_{ab}$
are odd.

The low energy dynamics of $Q_{ab}$ can be described by the
following Ginzburg-Landau field theory: \beqn \mathcal{L}_{Q} &=&
\mathrm{tr}[\partial_\mu Q^t \partial_\mu Q] + r (\mathrm{tr}[Q^t
Q]) + g (\mathrm{tr}[Q^t Q])^2 \cr\cr &+& u
(\mathrm{tr}[Q^tQQ^tQ]) + \cdots \label{LQ} \eeqn The first three
terms have an enlarged SO(9) symmetry which corresponds to the
rotation between the nine order parameters in matrix $Q$; while
the last term $\sim u$ breaks this SO(9) symmetry down to SO(4)
symmetry. Another term $\mathrm{Det}[Q]$ is also invariant under
SO(4) transformation, but $\mathrm{Det}[Q]$ breaks the
time-reversal and reflection symmetry of the honeycomb lattice,
therefore $\mathrm{Det}[Q]$ is forbidden in the Lagrangian
Eq.~\ref{LQ}. However, if the system already breaks the time
reversal and reflection symmetry (for instance $\langle
\bar{\zeta}\zeta \rangle \neq 0$), $\mathrm{Det}[Q]$ would be
allowed.

In Eq.~\ref{LQ}, when $r < 0$, $Q$ is ordered, and the SO(4)
symmetry is broken down to its subgroups. Depending on the sign of
$u$, there are two types of ground states:

Type $A$, $u < 0$, one example state of this phase is $\langle
Q_{33} \rangle \neq 0$, and all the other components $\langle
Q_{ab} \rangle = 0$. In this phase the SO(4) symmetry is broken
down to its following subgroup: \beqn \left( \ \mathrm{U(1)_{spin}
\otimes U(1)_{charge}} \otimes Z_2 \ \right)/Z_2. \eeqn The
$\mathrm{U(1)_{spin}}$ and $\mathrm{U(1)_{charge}}$ symmetry are
generated by matrices $S^z$ and $T^z$ in Eq.~\ref{algebra}. The
$Z_2$ in the numerator corresponds to reversing the direction of
$S^z$ and $T^z$ simultaneously, while keeping $Q_{33}$ invariant.
The $Z_2$ in the denominator is the same $Z_2$ as in
Eq.~\ref{so4}, which corresponds to changing the sign of electron
operator. The ground state manifold (GSM) of this phase is \beqn
\mathrm{GSM} \sim [S^2_{\mathrm{spin}} \otimes
S^2_{\mathrm{charge}}]/Z_2. \label{GSMA} \eeqn The ground state
can be described by two independent unit vectors $\vec{N}_s$ and
$\vec{N}_c$ which belong to the $(\bf{3},\bf{1})$ and
$(\bf{1},\bf{3})$ representation of SO(4) respectively, and
$Q_{ab} = N^a_c N^b_s$. This phase has four independent Goldstone
modes. The $Z_2$ in Eq.~\ref{GSMA} is due to the fact that
$\vec{N}_s$ and $\vec{N}_c$ can reverse direction simultaneously,
and the ground state remains invariant.

Type $B$, $u > 0$, one example state of this phase is $\langle
Q_{11} \rangle = \langle Q_{22} \rangle = \langle Q_{33}
\rangle\neq 0$. The SO(4) group element $G_{\mathrm{so(4)}}$ can
be written as $G_{\mathrm{su(2)},\mathrm{spin}} \otimes
G_{\mathrm{su(2)}, \mathrm{charge}}$, and the type $B$ phase
breaks the SO(4) down to its subgroup with
$G_{\mathrm{su(2)},\mathrm{spin}} = \pm G_{\mathrm{su(2)},
\mathrm{charge}}$. This implies that the residual symmetry group
elements can be parametrized as $\pm R^s(\theta^s, \vec{n}^s)$
with $\theta^s \in (0, 2\pi)$, which is equivalent to the diagonal
subgroup SU(2)$_+$ generated by operators $G^a = S^a + T^a $. Here
$R^s(\theta^s, \vec{n}^s)$ represents spin rotation by angle
$\theta^s$ about axis $\vec{n}^s$. The GSM of phase $B$ is \beqn
\mathrm{GSM} = \left( \ \mathrm{SO(4)} \otimes \mathcal{T} \
\right)/\mathrm{SU(2)}_+ = \mathrm{SO(3)} \otimes Z_2, \eeqn with
three Goldstone modes. $\mathcal{T}$ denotes the time reversal
symmetry, and type $B$ phase spontaneously breaks $\mathcal{T}$.
Therefore the GSM of phase $B$ contains two disconnected
sub-manifolds, with positive and negative $\mathrm{Det}[\langle Q
\rangle]$ respectively.

The order of $Q$ can be obtained through the following SO(4)
invariant interacting Lagrangian for Dirac fermions on the
honeycomb lattice: \beqn \mathcal{L} = \sum_{a = 1}^8
\bar{\zeta}_a \gamma_\mu\partial_\mu \zeta_a - g
\mathrm{tr}[Q^tQ], \label{diracinter}\eeqn The interaction $-
g\mathrm{tr}[Q^tQ]$ can be generated with SO(4) invariant
interaction on the lattice, for instance $\sum_{\ll i, j\gg}
\vec{S}_i \cdot \vec{S}_j \sim - \mathrm{tr}[Q^tQ] /8 + \cdots$. A
simple mean field calculation after the standard
Hubbard-Stratonovich Transformation of Eq.~\ref{diracinter} shows
that the type $A$ phase has more favorable ground state energy
compared with the type $B$ phase on the honeycomb lattice. In the
following we will mainly focus on the analysis on the type $A$
phase.

\section{Phase diagram around Type A phase} \label{A}

Now we hope to understand the topological defects, and the phase
transitions driven by topological defects in phase $A$. We will
first give an argument about the phase diagram and phase
transitions using the quantum numbers carried by the topological
defects, and then a systematic description based on the Majorana
liquid formalism developed in Ref.~\cite{xs2010} will be
presented. We will demonstrate that these two approaches match
very well.

\subsection{Topological defects and phase transitions} \label{GLPDA}

In phase $A$, since the GSM is $[S^2 \otimes S^2]/Z_2$, both spin
and charge sectors can have Skyrmion like defects characterized by
homotopy group $\pi_2[S^2]$. Again, let us assume that $Q_{33}$ is
the only component that acquires a nonzero expectation value, then
$\langle Q_{33} \rangle$ breaks the SO(4) symmetry down to
residual symmetries generated by $S^z$ and $T^z$ in
Eq.~\ref{algebra}. According to
Ref.~\cite{senthil2007,abanov2000,abanov2001}, under our current
assumption that $\vec{N}_c \ \| \ \hat{z} $ ($Q_{ab}$ is the QSH
vector), a Skyrmion of the spin sector manifold
$S^2_{\mathrm{spin}}$ carries charge $2e$; and a Skyrmion current
is identified as the charge current: \beqn J^{\mu} &=&
\frac{2e}{8\pi} \epsilon_{\mu\nu\rho} \int d^2x \epsilon_{abc}
\hat{N}^a_s
\partial_\nu \hat{N}^b_s \partial_\rho \hat{N}^c_s. \eeqn For the same
reason, a Skyrmion of the charge sector manifold will carry
spin-1: $S^z = 1$. For a general state with $\langle Q_{ab}
\rangle \neq 0$, the spin-Skyrmion carries the quantum number of
the $\mathrm{U(1)_{charge}}$ residual symmetry, while the
charge-Skyrmion carries $\mathrm{U(1)_{spin}}$ quantum number
$i.e.$ spin and charge view each other as topological defects.

The condensation of Skyrmions with nontrivial quantum numbers can
lead to unconventional quantum phase transitions. For instance the
proposal of deconfined quantum criticality is based on the
observation that the Skyrmion of the N\'{e}el order carries
lattice momentum \cite{sachdev1990,senthil2004a,senthil2004},
hence the condensate of the Skyrmion is equivalent to the valence
bond solid state. In our current case, since a charge-Skyrmion
carries spin-$1$, if the charge-Skyrmion is condensed, then the
$\mathrm{SU(2)_{charge}}$ symmetry is fully restored, which
implies that the condensate of the charge Skyrmion is a Mott
insulator. Meanwhile, the residual spin symmetry is further
spontaneously broken down to $Z_{2} \otimes Z_2$. One of these
$Z_2$ corresponds to changing the sign of electron, the other one
corresponds to reversing the direction of $\vec{N}_s$. The GSM of
the charge Skyrmion condensate is \beqn \mathrm{GSM} =
\mathrm{SU(2)_{spin}}/[Z_2 \otimes Z_2] = \mathrm{SO(3)}/Z_2.
\eeqn We mod $Z_2$ from SO(3), because after the proliferation of
charge Skyrmion, $\vec{N}_c$ is completely disordered, and
$\vec{N}_s$ becomes a headless vector, due to the $Z_2$ in
Eq.~\ref{GSMA}.

How do we determine the order of the charge-Skyrmion condensate
unambiguously? As was pointed out in Ref.~\cite{senthil2006}, the
phase of the O(3) Skyrmion condensate can be identified as order
parameters that share an O(5) Wess-Zumino-Witten (WZW) term with
the O(3) order parameter. Therefore, to unambiguously identify the
order of spin-Skyrmion condensate, we need to seek for order
parameters that have an O(5) WZW term with vector $\varphi^a =
Q_{ab}$. It turns out that the N\'{e}el order parameter $\vec{N}$
is the only candidate of the charge-Skyrmion condensate. For
arbitrary $b$, we obtain the following WZW term between $\varphi^a
= Q_{ab}$ and $\vec{N} \sim \bar{\chi} \mu^y \vec{S}\chi$: \beqn
\mathcal{L} &=& \sum_{a = 1}^5 \frac{1}{g}(\partial_\mu \phi^a)^2
\cr\cr &-& \frac{3i}{4\pi}\int du d^3x \epsilon_{abcde}\phi^a
\partial_x \phi^b \partial_y \phi^c \partial_\tau \phi^d \partial_u \phi^e, \cr\cr
&& \phi^{a} = \varphi^{a} = \bar{\chi}T^{a} S^b\chi , \ \ a = 1,
2, 3, \cr \cr && \phi^4 = N^c \sim \bar{\chi} \mu^y S^c \chi ,
\cr\cr && \phi^5 = N^d \sim \bar{\chi}\mu^y S^d\chi, \ \ c, d \neq
b. \label{wzw}\eeqn Therefore the charge-Skyrmion condensate
contains both headless vector $\vec{N}_s$ and N\'{e}el order
$\vec{N}$, and $\vec{N}_s \perp \vec{N}$. Physically the headless
vector $\vec{N}_s$ corresponds to spin nematic order
$\mathcal{S}_{ab} = 3N^a_sN^b_s - \delta^{ab} (\vec{N}_s)^2 $,
which is invariant under reversing the direction of $\vec{N}_s$.
All these results will be confirmed later with the Majorana liquid
formalism.

Notice that manifold SO(3) is equivalent to the projected manifold
$S^3/Z_2$, which gives us a convenient way of parameterizing
SO(3). Let us introduce SU(2) spinon $z_\alpha$ with constraint
$|z_1|^2 + |z_2|^2 = 1$. This constraint implies that the SU(2)
spinon $z_\alpha$ parametrizes $S^3$. Then by coupling $z_\alpha$
to a $Z_2$ gauge field, the gauge invariant GSM of the condensate
of $z_\alpha$ automatically becomes SO(3) \cite{senthil1994}.
SO(3) manifold can also be viewed as the manifold of all the
configurations of three perpendicular unit vectors $\vec{q}_1$,
$\vec{q}_2$ and $\vec{q}_3$. These three vectors can be
parametrized as \beqn \vec{q}_1 = z^\dagger \sigma^b z, \ \
\vec{q}_2 + i\vec{q}_3 = z^t i \sigma^y\sigma^b z,
\label{cpz2}\eeqn which automatically guarantees the
perpendicularity of these vectors. In our situation, the three
perpendicular vectors that characterize the GSM are $\vec{N}_s$,
$\vec{N}$ and $\vec{N}_s\times \vec{N}$. Since $\vec{N}_s$ is
headless, the GSM is in fact $\mathrm{SO(3)}/Z_2$. And in the next
subsection we will demonstrate that it is most convenient to
describe this GSM by introducing a $Z_2 \otimes Z_2$ or $Z_4$
gauge field.

Manifold SO(3) has homotopy group $\pi_1[\mathrm{SO(3)}] = Z_2$,
therefore phase $B$ has topologically stable half vortex. Using
the CP(1) spinon description introduced in Eq.~\ref{cpz2}, this
half-vortex can also be viewed as the vison (a dynamical
$\pi$-flux) of the $Z_2$ gauge field coupled to $z_\alpha$.
Pictorially, a vison can be viewed as a configuration with (for
instance) $\vec{q}_1$ being uniform in space, while $\vec{q}_2$
and $\vec{q}_3$ have a vortex. Now since $\vec{N}_s$ and
$\vec{N}_s \times \vec{N}$ are both headless vectors, this state
also supports ``half vison", where $\vec{N}$ is uniform, while
$\vec{N}_s$ has a half vortex in space. In fact, this half vison
has a counterpart in phase $A$. Since phase $A$ has GSM $[S^2
\otimes S^2]/Z_2$, there exists ``double half vortex", and both
$\vec{N}_s$ and $\vec{N}_c$ reverse direction after encircling
this double half vortex. After phase $A$ is destroyed by
proliferating the charge-Skyrmion, this double half vortex becomes
the half vison of $\mathrm{SO(3)}/Z_2$.

In the spin-charge dual side of the theory, all the conclusions
can be obtained by straightforward generalization. Once the
spin-Skyrmion is condensed, the system will also enter a phase
with GSM $\mathrm{SO(3)}/Z_2$, and the $\mathrm{SU(2)_{spin}}$
symmetry is fully restored, which implies that the condensate of
the spin-Skyrmion is spin singlet. In Ref.~\cite{senthil2007}, the
authors proposed that after the proliferation of Skyrmions of the
QSH vector, the system enters a spin singlet $s-$wave
superconductor. In our situation, since there is a generic
$\mathrm{SU(2)_{charge}}$ symmetry, the $s-$wave superconductor is
promoted to a phase with GSM $\mathrm{SO(3)}/Z_2$. If $\vec{N}_c \
\| \ \hat{z}$ ($Q_{ab}$ is the QSH vector), the
$\mathrm{SO(3)}/Z_2$ manifold is characterized with headless
vector $\vec{N}_c$ and $s-$wave superconductor. The order of
headless vector $\vec{N}_c$ implies that the degeneracy between
CDW and $s-$wave superconductor is spontaneously lifted. In
general, the order after spin-Skyrmion proliferation can also be
determined with the same WZW term analysis in Eq.~\ref{wzw}. A
full list of order parameters with WZW terms can be found in
Ref.~\cite{ryuwzw}.

This Skyrmion condensation transition is described by the same
CP(1) field theory as the deconfined quantum criticality
\cite{senthil2004,senthil2004a}. How do we see the CP(1)
transition directly? The CP(1) model $\mathcal{L} =
\frac{1}{g}|(\partial_\mu - iA_\mu)z|^2$ describes a transition
between a condensate of spinon $z_\alpha$ and a photon phase. The
spinon condensate has GSM $S^2$, while the photon phase has GSM
$S^1$, as it is a condensate of the U(1) gauge flux. In our case
the Skyrmion condensation is a transition between $[S^2\otimes
S^2]/Z_2$ and $\mathrm{SO(3)}/Z_2$, while SO(3) can be roughly
viewed as $S^2 \otimes S^1$, therefore effectively the Skyrmion
condensation is more or less also a transition between $S^2$ and
$S^1$, so it is equivalent to the CP(1) transition. This
hand-waving argument will be made precise in the next subsection
by the Majorana liquid formalism.

Since a spin-Skyrmion (charge-Skyrmion) carries charge-2$e$
(spin-1), then the corresponding half-Skyrmion (vortex) will carry
charge-$e$ and spin-1/2 respectively. If a charge-$e$ excitation
encircles around a spin-1/2 excitation bound with a charge-vortex,
the charge-$e$ excitation will acquire a $\pi$ phase shift; on the
other hand, if a spin-1/2 excitation encircles a charge-$e$
excitation bound with a spin-vortex, the spin-1/2 excitation will
also gain a $\pi$ Berry phase. This implies that in phase $A$
charge-$e$ and spin-1/2 excitations have mutual semion statistics.

In phase $A$, a vortex is not a local excitation, and the gapless
Goldstone mode of phase $A$ makes the adiabatic braiding between
two excitations impossible, therefore the semion statistics in
phase $A$ is not well defined. However, later we will see that
phase $A$ is adjacent to a liquid phase where the spin-charge
mutual semion statistics persists, and it becomes a well defined
property.

\begin{figure}
\includegraphics[width=2.6 in]{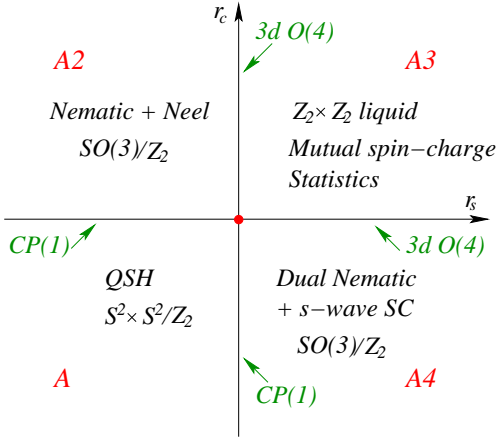}
\caption{Phase diagram around type $A$ phase with $\langle Q_{33}
\rangle \neq 0$.} \label{pd}
\end{figure}

\subsection{Phase diagram with Majorana liquid formalism}

From now on we hope to understand the phase diagrams discussed
above with a more solid formalism. In Ref.~\cite{xs2010}, we
discussed a fractionalized phase of electrons by decomposing
$\zeta$ as following: \beqn \zeta &=& Z_s Z_c \ \chi, \cr\cr Z_s
&=& \phi^s_0 + i \phi^s_1 S^x + i \phi^s_2 S^y + i \phi^s_3 S^z,
\cr \cr Z_c &=& \phi^c_0 + i \phi^c_1 T^x + i \phi^c_2 T^y + i
\phi^c_3 T^z . \label{Rz} \eeqn The electron $\zeta$ decomposes
into the bosonic fields $Z_s$ and $Z_c$ carrying its spin and
charge respectively, and into the Majorana fermion $\chi$ carrying
the Fermi statistics. The resulting theory has a ${\rm SO}(4)_g =
{\rm SU}(2)_{s,g} \otimes {\rm SU}(2)_{c,g}$ gauge invariance:
$Z_s$ and $\chi$ carry SU(2)$_{s,g}$ charges, and $Z_c$ and $\chi$
carry SU(2)$_{c,g}$ charges.

After the operator decomposition, when both $Z_s$ and $Z_c$ are
gapped out, one obtains the parent state, $i.e.$ the Algebraic
Majorana Liquid (AML) state with Lagrangian \beqn \mathcal{L}_{\rm
AML} &=& \bar{\chi} \gamma_\mu \left(
\partial_\mu - i A^a_{s,\mu} S^a - i A^a_{c,\mu}
T^a \right) \chi. \label{aml} \eeqn The fractionalized Majorana
fermion $\chi$ fills the same mean field band structure as the
physical Majorana fermion $\zeta$. The gauge field $A^a_{s,\mu}$
and $A^a_{c,\mu}$ also couple to the spin and charge SU(2) rotors
$Z_s$ and $Z_c$ as well. Since $\chi$ no longer carries physical
spin and charge quantum numbers, the fermion bilinears of $\chi$
can only break the gauge symmetry, but not physical symmetry. If
$Z_s$ or $Z_c$ condense, the formalism reduces to the two standard
slave particle formalisms, with fermionic \cite{wen2002a} or
bosonic spinons \cite{sachdev1990} respectively.

Now let us assume $\chi$ enters a type $A$ phase $i.e.$ the matrix
field $\tilde{Q}_{ab} = \bar{\chi} A_{ab} \chi $ condenses. For
instance let us take \beqn \langle \tilde{Q}_{33} \rangle =
\langle \bar{\chi} \sigma^z \chi \rangle \neq 0. \eeqn Although
the fractionalized Majorana fermion $\chi$ fills the same mean
field band structure as $\zeta$, unlike the physical QSH vector,
nonzero $\langle \tilde{Q}_{3b} \rangle $ breaks no discrete
symmetries (time-reversal, refection) when rotor fields $Z_s$ and
$Z_c$ are gapped. This is because the gauge symmetry released from
gapping out the rotor fields can always reverse the sign of
$\langle \tilde{Q}_{3b} \rangle $. This condensate of
$\tilde{Q}_{33}$ breaks the $\mathrm{SU(2)}_{s,g} \otimes
\mathrm{SU(2)}_{c,g}$ gauge invariance down to
$\mathrm{U(1)}_{s,g} \otimes \mathrm{U(1)}_{c,g}$ gauge symmetry
generated by $S^z$ and $T^z$. Sometimes it will be convenient to
use the following spin and charge CP(1) field \beqn z^{s} =
(z^s_1, \ z^s_2)^t = (\phi^s_0 + i\phi^s_3, \ -\phi^s_2 + i
\phi^s_1)^t, \cr\cr z^{c} = (z^c_1, \ z^c_2)^t = (\phi^c_0 +
i\phi^c_3, \ -\phi^c_2 + i \phi^c_1)^t. \label{cp1}\eeqn It was
discussed in our previous work \cite{xs2010} that, after
integrating out $\chi$, we obtain the low energy theory when
$\langle \tilde{Q}_{33} \rangle \neq 0$, which is a mutual
Chern-Simons theory: \beqn \mathcal{L}_{cs} &=&
\frac{2i}{2\pi}\epsilon_{\mu\nu\rho} A^z_{c,\mu}\partial_\nu
A^z_{s,\rho} \cr\cr &+& |(\partial_\mu - i A^z_{s,
\mu})z^{s}_\alpha|^2 + r_s|z^{s}_\alpha|^2 \cr\cr &+&
|(\partial_\mu - i A^z_{c, \mu})z^{c}_\alpha|^2 +
r_c|z^{c}_\alpha|^2 + \cdots \label{lcs} \eeqn The CP(1) fields
$z^{s}_\alpha$ and $z^{c}_\alpha$ carry spin and charge
respectively, and we have chosen the notation to make both spin
and charge SU(2) physical global symmetries manifest.
%$i.e.$ the
%physical symmetries are just SU(2) symmetries on CP(1) fields
%$z^s$ and $z^c$.
Eq.~\ref{lcs} implies that there is a mutual semion statistics
between the charge and spin CP(1) fields $z^c_\alpha$ and
$z^s_\alpha$, which verifies the observation in section
\ref{GLPDA}.

This field theory is similar to the one on the triangular lattice
\cite{xs}, with mutual semion statistics between charge and vison,
except there the SU(2) symmetry of the vison is broken down to
discrete symmetry by higher order terms \cite{sondhi2001a}, while
here the SU(2) charge symmetry is exact. Based on this analogy, we
can propose a similar global phase diagram with tuning parameters
$r_s$ and $r_c$ (Fig.~\ref{pd}) as Ref.~\cite{xs}, with a
different interpretation of the phases:

\subsubsection{Phase $A$}

Phase $A$ is the phase with both $z^s_\alpha$ and $z^c_\alpha$
condensed, and the $\mathrm{SU(2)}_{s,g}$ and
$\mathrm{SU(2)}_{c,g}$ gauge fields are both Higgsed and gapped
out from the spectrum. This phase is characterized by the SU(2)
vectors: \beqn N^a_s = z^{s \dagger}\sigma^a z^s \sim
\mathrm{tr}[Z^t_s S^a Z_s S^z]; \cr\cr N^a_c = z^{c
\dagger}\sigma^a z^c \sim \mathrm{tr}[Z^t_c T^a Z_c T^z],
\label{vectors}\eeqn and the gauge invariant physical order
parameter is \beqn Q_{ab} &=& \bar{\zeta} A_{ab} \zeta =
\bar{\chi} Z_s^t Z_c^t A_{ab}Z_sZ_c \chi \cr\cr  &\sim&
\mathrm{tr}[Z^t_c T^a Z_c T^f] \ \mathrm{tr}[Z^t_s S^b Z_s S^g] \
\langle \bar{\chi} A_{fg} \chi\rangle \cr\cr &\sim& \langle
\tilde{Q}_{33}\rangle N^a_sN^b_c, \label{Qfrac}\eeqn therefore the
physical GSM is $[S^2 \otimes S^2]/Z_2$. This phase is precisely
the phase $A$ obtained in the GL formalism in section \ref{secGL}.
The Skyrmion of vector $\vec{N}_s$ is equivalent to the flux of
gauge field $A^z_{s,\mu}$, and due to the mutual CS interaction,
the gauge flux of $A^z_{s,\mu}$ (Skyrmion of $\vec{N}_s$) carries
charge excitation $z^c_\alpha$, which confirms our analysis in
section \ref{GLPDA}.

\subsubsection{Phase $A2$ and $A4$} \label{GSMA2}

Phase $A2$ has $r_s < 0$, $r_c > 0$, hence it is a phase with
$z^s_\alpha$ condensed while $z^c_\alpha $ gapped. In section
\ref{GLPDA} we concluded that the GSM of this phase is
$\mathrm{SO(3)}/Z_2$, and it has both nematic order and N\'{e}el
order by directly calculating the WZW term. How do we understand
the physical orders using the Majorana liquid formalism? Since
$z^c_\alpha$ is gapped, in phase $A2$ there is no charge degrees
of freedom, therefore the ``QSH" vector $\tilde{Q}_{3b}$ should
correspond to a pure spin operator. In fact, since $z^c_\alpha$ is
gapped, $\langle Q_{ab} \rangle $ in Eq.~\ref{Qfrac} vanishes,
hence the only gauge invariant operator which acquires a nonzero
expectation value is \beqn && 3\tilde{Q}^t_{cd}\tilde{Q}_{de} \
\mathrm{tr}[Z_s^tS^aZ_sS^c] \ \mathrm{tr}[Z_s^tS^bZ_sS^e] \cr\cr
&& \sim \langle \tilde{Q}_{33} \rangle^2 N^a_sN^b_s,
\label{nematic2}\eeqn therefore we can define the physical order
parameter as \beqn \mathcal{S}_{ab} \sim 3N^a_sN^b_s -
\delta^{ab}(\vec{N}_s)^2 \sim 3S^a_iS^b_j - \delta^{ab} \vec{S}_i
\cdot \vec{S}_j. \label{nematic}\eeqn Hence $\mathcal{S}_{ab}$ is
the spin-2 nematic order parameter which breaks the spin rotation
symmetry down to $\mathrm{U(1)} \otimes Z_2$, but preserves the
discrete symmetries. Notice that the physical order parameter is
always a bilinear of $\vec{N}_s$.

When $z^s_\alpha$ is condensed, the $\mathrm{SU(2)}_{s,g}$ gauge
field is Higgsed, then the $\mathrm{SU(2)}_{s,g}$ gauge charge of
$\chi$ becomes equivalent to the physical spin quantum number of
$\zeta$ after a SU(2) gauge transformation. If we take $\langle
\tilde{Q}_{33} \rangle \neq 0$, the low energy field theory for
fermions in phase $A2$ reads: \beqn \mathcal{L} &=&
\bar{\psi}\gamma_\mu(\partial_\mu - i A^z_{c,\mu})\psi + m \langle
\tilde{Q}_{33} \rangle \cdot \bar{\psi}\sigma^z \psi,
\label{qshxy}\eeqn where $\psi = \chi_1 + i\chi_2$. Based on the
QSH physics, the flux of gauge field $A^z_{c,\mu}$ carries spin:
\beqn \nabla \times \vec{A}^z_{c} \sim \psi^\dagger \sigma^z \psi
\ \mathrm{tr}[Z_s^t S^a Z_s S^z]. \eeqn This equivalence between
the flux and spin implies that the photon phase of gauge field
$A^z_{c,\mu}$, which is the condensate of the flux is a spin XY
order. This effect was studied in Ref.~\cite{ran2007,ran2008} with
projected wave-function calculation, and the photon phase of the
U(1) gauge field is precisely the N\'{e}el order: \beqn N^a \sim
\bar{\chi} \mu^y S^c \chi \ \mathrm{tr}[Z_s^t S^a Z_s S^c].
\label{neelfra}\eeqn Based on these analysis, we conclude that
phase $A2$ is a phase with both nematic vector $\vec{N}_s$ in
Eq.~\ref{vectors} and AF N\'{e}el order $\vec{N}$ in
Eq.~\ref{neelfra}. Eq.~\ref{vectors} and Eq.~\ref{neelfra}
guarantee that $\vec{N}_s \perp \vec{N}$: \beqn \vec{N}_s \cdot
\vec{N} \sim \sum_a \mathrm{tr}[Z^t_s S^a Z_s S^z] \
\mathrm{tr}[Z^t_s S^a Z_s S^x] = 0. \eeqn As we mentioned before,
since the nematic vector is headless, the GSM should be
$\mathrm{SO(3)}/Z_2$. This analysis again confirms our prediction
in section \ref{GLPDA} with the WZW term.

Phase $A4$ is the spin-charge dual phase of phase $A2$, the GSM is
also $\mathrm{SO(3)}/Z_2$ with three branches of Goldstone modes.
The spin-charge dual of the inplane N\'{e}el order is precisely a
$s-$wave superconductor. The spin-charge dual of the nematic order
$\mathcal{S}_{ab}$ will break the $\mathrm{SU(2)_{charge}}$, for
instance \beqn \mathcal{S}_{zz} \sim 2 T_i^z T_j^z - T^x_i T^x_j -
T^y_i T^y_j, \label{chargenematic} \eeqn with $\vec{T}_i$ given by
Eq.~\ref{hubbard}. Therefore, if we turn on an extra density
repulsion between next nearest neighbor sites in
Eq.~\ref{hubbard}, it corresponds to turning on
$\mathcal{S}_{zz}$, and breaks the $\mathrm{SU(2)_{charge}}$ down
to $\mathrm{U(1)}\otimes Z_2 $. This U(1) corresponds to the
ordinary electron charge conservation, and $Z_2$ corresponds to
the discrete particle-hole symmetry.

\subsubsection{Phase $A3$}

Phase $A3$ is a liquid state with $r_s
> 0$, $r_c > 0$, both $z^s_\alpha$ and $z^c_\alpha$
are gapped out. When $z^c_\alpha$ is gapped, $A^z_{c,\mu}$ is in
the photon phase. Since the photon phase of 2+1d U(1) gauge field
is also the condensate of gauge flux based on the standard
QED-superfluid duality, the mutual CS coupling in Eq.~\ref{lcs}
implies that the photon phase of $A^z_{c,\mu}$ breaks the
$\mathrm{U(1)}_{s,g}$ down to $Z_2$ gauge symmetry. For the same
reason, $\mathrm{U(1)}_{c,g}$ is also broken down to $Z_2$. The
mutual CS theory in Eq.~\ref{lcs} has the same topological
degeneracy as the standard $Z_2$ gauge field on the torus
\cite{xs,koulevin}, and the mutual statistics between charge and
spin is an analogue of the mutual statistics between charge and
vison of the well-known toric code model \cite{kitaev2003}.

In addition to the $Z_2$ gauge field coming from the mutual CS
coupling, there is one extra residual $Z_2$ gauge symmetry which
corresponds to reversing the sign of gauge symmetry generators
$S^z$ and $T^z$ simultaneously, while leaving $\tilde{Q}_{33}$
invariant. This extra discrete $Z_2$ gauge symmetry contains group
elements \beqn G^{(z2)} =  I_{4\times 4}, \ \mathrm{or} \ \
S^xT^x. \eeqn This $Z_2$ gauge field couples to both spin and
charge SU(2) rotors $Z_s$ and $Z_c$, but it was not explicit in
our continuum limit field theory. Therefore phase $A3$ is
characterized by $Z_2 \otimes Z_2$ gauge fields. Under $Z_2$ gauge
symmetry $G^{(z2)}$, the fractionalized particles transform as
\beqn Z_{s,j} &\rightarrow& Z_{s,j} \frac{1}{2}\left( (1 +
\mathcal{\mu}_{j})I_{4\times 4} + (1 - \mathcal{\mu}_{j})i
S^x\right), \cr\cr Z_{c,j} &\rightarrow& Z_{c,j} \frac{1}{2}\left(
(1 + \mathcal{\mu}_{j})I_{4\times 4} + (1 - \mathcal{\mu}_{j})i
T^x\right), \cr\cr \chi_{j} &\rightarrow& \frac{1}{2}\left( (1 +
\mathcal{\mu}_{j})I_{4\times 4} + (1 - \mathcal{\mu}_{j}) S^x
T^x\right)\chi_j, \cr\cr && \mu_j = \pm 1. \eeqn If the matter
fields are ignored, these two $Z_2$ gauge fields are equivalent to
a $Z_4$ gauge field with group elements $G^{(z4)} = \exp[i \theta
S^xT^x]$, $\theta = 0, \ \pi/2, \ \pi, \ 3\pi/2$.

These two $Z_2$ gauge field together again implies that the GSM of
$A2$ (the condensate of $z^s_\alpha$) is $\mathrm{SO(3)}/Z_2$, as
we already concluded. If we approach phase $A2$ from phase $A3$,
we can interpret phase $A2$ as the condensate of SU(2) spin rotor
$Z_s$ which couples to the two $Z_2$ gauge groups discussed above.
With the condensate of $Z_s$ we can again define three
perpendicular vectors: \beqn \vec{N}_s & = & \mathrm{tr}[Z_s^t
\vec{S} Z_s S^z] \sim z^{s\dagger}\vec{\sigma} z^s, \cr\cr
\vec{N}_1 & = & \mathrm{tr}[Z_s^t \vec{S} Z_s S^x] \sim
\mathrm{Re}[(z^{s})^t i\sigma^y \vec{\sigma} z^s], \cr\cr
\vec{N}_2 & = & \mathrm{tr}[Z_s^t \vec{S} Z_s S^y] \sim
\mathrm{Im}[(z^{s})^t i\sigma^y \vec{\sigma} z^s]. \eeqn
$\vec{N}_s$ and $\vec{N}_2$ change sign under gauge transformation
$Z_s \rightarrow Z_s iS^x$ ($i.e. \ \mu_j = -1$), while
$\vec{N}_1$ never changes sign, therefore $\vec{N}_s$ and
$\vec{N}_2$ become headless nematic vectors by coupling to the
$Z_2 \otimes Z_2$ gauge group. Hence the manifold formed with
$\vec{N}_s$, $\vec{N}_1$ and $\vec{N}_2$ is $\mathrm{SO(3)}/Z_2$.
This is completely consistent with our description of phase $A2$
in section \ref{GLPDA} and \ref{GSMA2}.

In addition to the mutual statistics between $Z_s$ and $Z_c$,
there is one more topological defect in phase $A3$ with \beqn
\prod_{\mathcal{C}} G^{(z2)} = S^xT^x, \eeqn $\mathcal{C}$ is a
closed loop on the lattice. This defect carries a gauge flux
$S^xT^s$. After encircling this defect, $Z_s \rightarrow Z_s
iS^x$, $Z_c \rightarrow Z_c iT^x$. The vectors $\vec{N}_s$ and
$\vec{N}_2$ always acquire a minus sign after encircling this
defect. Therefore this defect is a counterpart of the ``double
half vortex" and ``half vison" discussed in section \ref{GLPDA}.

\subsubsection{Discussion}

The universality class of the phase transitions in Fig.~\ref{pd}
can also be analyzed with field theory Eq.~\ref{lcs}, in the same
way as Ref.~\cite{xs}. Quoting the results in Ref.~\cite{xs}, the
transition between phases $A3$, $A2$, and the transition between
phases $A3$, $A4$ are 3d O(4) transitions, because spinon
$z^s_\alpha$ and $z^c_\alpha$ are O(4) vectors, and the fully
gapped discrete gauge fields coupled to the O(4) vector do not
change the O(4) universality class \cite{senthil1994}. The
transition ($A$, $A3$), and the transition ($A$, $A4$) are CP(1)
transitions, which become manifest with the CP(1) fields
$z^s_\alpha$ and $z^c_\alpha$, and U(1) gauge fields
$A^z_{s,\mu}$, $A^z_{c,\mu}$.

A recent quantum Monte carlo simulation of the Hubbard model on
the honeycomb lattice suggests that there is a fully gapped spin
liquid phase \cite{meng} sandwiched between the ordinary N\'{e}el
order and semi-metal phase, which has motivated spin liquid
analysis on the honeycomb lattice using either slave boson or
slave fermion techniques \cite{wang2010,ran2010b}. In our
formalism, phase $A3$ in phase diagram Fig.~\ref{pd} is a
candidate of this gapped spin liquid. However, based on our
analysis, phase $A3$ is not directly adjacent to a pure N\'{e}el
order, instead phase $A3$ is adjacent to phase $A2$ with both
N\'{e}el order and nematic order. Starting with phase $A2$, we
need to go through one more transition which suppresses the
nematic order, and enters the final N\'{e}el order in the large
Hubbard $U$ limit.

The spin-2 nematic order is a natural candidate of the ground
state of spin-1 systems, with bi-quadratic interactions
\cite{balentssimon}. For spin-1/2 models, nematic order can exist
when there is a considerable ring exchange or multi-spin
interaction, which can be generated in the weak Mott insulator
phase of the simplest Hubbard model with high order perturbation
of $t/U$. Our prediction of a phase with coexistence of nematic
and N\'{e}el order can be checked numerically in future. We will
present a general classification about nematic orders and their
adjacent spin liquid phases in future \cite{xufuture}.

A similar analysis can be applied to the $Z_2$ liquid phase
obtained from the standard Schwinger boson formalism. Since the
spinon $z_\alpha$ always couples to a $Z_2$ gauge field, the
condensate of $z_\alpha$ is not an ordinary N\'{e}el order,
because there always exists three perpendicular gauge invariant
vectors like Eq.~\ref{cpz2}. In Ref.~\cite{ran2010}, the authors
proposed that there is an intermediate chiral antiferromagnetic
order between a fully gapped $Z_2$ liquid phase and a N\'{e}el
order. This chiral AF state has GSM SO(3), which is different from
the phase predicted in our paper with both nematic and N\'{e}el
order.

\section{Phase diagram around Type B phase} \label{B}

Now let us move on the phase $B$ with GSM $\mathrm{SO(3)}\otimes
Z_2$. All the phases discussed in this section breaks
time-reversal symmetry $\mathcal{T}$, therefore we only focus on
one of the two disconnected sub-manifolds SO(3). Phases with GSM
SO(3) have been studied extensively with non-collinear spin
density wave \cite{senthil1994}. After disordering the state, both
$\mathrm{SU(2)_{spin}}$ and $\mathrm{SU(2)_{charge}}$ are
restored, while the vison of the SO(3) manifold is still locally
conserved, and the system most naturally enters a $Z_2$ liquid
phase. %The phase transition between phase $B$ and and the $Z_2$
%liquid phase is again a 3d O(4) transition \cite{senthil1994}.

\begin{figure}
\includegraphics[width=2.6 in]{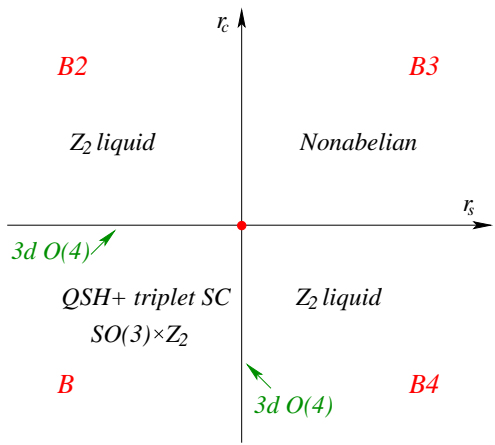}
\caption{Phase diagram around type $B$ phase with $\langle Q_{11}
\rangle = \langle Q_{22} \rangle = \langle Q_{33} \rangle \neq
0$.} \label{pd2}
\end{figure}

Again, we hope to understand the phase diagram with the Majorana
liquid formalism. Let us assume $\chi$ in the parent state
Eq.~\ref{aml} enters the type $B$ phase, for instance
$\langle\tilde{Q}_{11}\rangle = \langle\tilde{Q}_{22}\rangle =
\langle\tilde{Q}_{33}\rangle \neq 0$. It would be convenient to
introduce the following CP(1) fields \beqn z^{s} = (z^s_1, \
z^s_2)^t = (\phi^s_0 - i\phi^s_3, \ \phi^s_2 - i \phi^s_1)^t,
\cr\cr z^{c} = (z^c_1, \ z^c_2)^t = (\phi^c_0 - i\phi^c_3, \
\phi^c_2 - i \phi^c_1)^t. \eeqn The phase diagram around type $B$
order is depicted in Fig.~\ref{pd2}:

Phase $B$ in Fig.~\ref{pd2} with both $z^s_\alpha$ and
$z^c_\alpha$ condensed is precisely the phase $B$ in the
Ginzburg-Landau description in section \ref{secGL}, with GSM
$\mathrm{SO(3)}\otimes Z_2$. In phase $B2$, the
$\mathrm{SU(2)}_{s,g}$ gauge symmetry is Higgsed by the
condensation of $z^s_\alpha$, while the $\mathrm{SU(2)}_{c,g}$ is
broken down to $Z_2$ gauge symmetry by $\langle \tilde{Q}_{ab}
\rangle$, this is the $Z_2$ liquid phase we discussed above. The
residual $Z_2$ gauge symmetry is the subgroup of
$\mathrm{SU(2)}_{c,g}$ that changes the sign of $\chi$. Phase $B4$
is the same $Z_2$ liquid phase as phase $B2$.

%The phase transition ($B$, $B2$) and the transition ($B$, $B4$)
%are described by the condensation of SU(2) rotor $z^s_\alpha$ or
%$z^c_\alpha$ coupled with $Z_2$ gauge field, therefore again they
%belong to the 3d O(4) universality class.

Now we turn to phase $B3$ in Fig.~\ref{pd}. In this phase both
$z^s_\alpha$ and $z^c_\alpha$ are gapped, while $\langle
\tilde{Q}_{ab} \rangle$ breaks the SO(4) gauge symmetry down to
$\mathrm{SU(2)_+} = \mathrm{SU(2)}_{s,g} + \mathrm{SU(2)}_{c,g}$,
hence in this phase there is only one SU(2) gauge field $A^a_\mu
G^a$, with $G^a = S^a + T^a$. After integrating out the fermion
$\chi$, the following Chern-Simons term is induced for the
residual $\mathrm{SU(2)}_+$ gauge field: \beqn \mathcal{L} &=&
\frac{2}{4\pi} \mathrm{tr}\left( A \wedge dA + \frac{2}{3} A
\wedge A \wedge A \right) \cr\cr &+& |(\partial_\mu - \sum_a
A^a_\mu \sigma^a)z^s|^2 + r_s|z^s|^2 \cr\cr &+& |(\partial_\mu -
\sum_a A^a_\mu \sigma^a)z^c|^2 + r_c|z^c|^2 + \cdots \cr\cr A &=&
A^a_\mu\sigma^a dx^\mu, \label{su2cs}\eeqn therefore phase $B3$ is
characterized by SU(2) CS theory at level 2, which is a nonabelian
theory \cite{chetan2009}.

A different way of obtaining the same theory, is by turning on
another order parameter $\bar{\chi}\chi$ in addition to
$\tilde{Q}_{ab}$. The order parameter $\bar{\chi}\chi$ will drive
the system into a quantum Hall state, and lead to the SU(2)$_1$ CS
theory for both $A^a_{s,\mu} $ and $A^a_{c,\mu}$, which is similar
to the CS effective theory of the chiral spin liquid state
\cite{wen2002a}. The order $\langle \tilde{Q}_{ab} \rangle$
requires $A^a_{s, \mu} = A^a_{c,\mu} = A^a_\mu$, therefore the
final theory becomes the SU(2) CS theory at level 2 in
Eq.~\ref{su2cs}. Now by reducing the order $\langle \bar{\chi}\chi
\rangle$ to zero, the SU(2) CS theory is unchanged. More detailed
properties of this phase will be further discussed in future
\cite{xufuture}.

In phase $B3$, the residual gauge symmetry $G_{s,g}$ is either
$G_{c,g}$ or $- G_{c,g}$. This $Z_2$ structure does not show up in
the Lie Algebra of the gauge group, but it implies that there is
one extra $Z_2$ gauge field that couples to $\chi$ and either one
of $z^s_\alpha$ or $z^c_\alpha$. Hence when $z^s_\alpha$ or
$z^c_\alpha$ condenses, the gauge field $A^a_\mu$ is Higgsed, but
the system still has a $Z_2$ gauge symmetry, which characterizes
the $Z_2$ liquid phase $B2$ and $B4$.

The transition ($B2$, $B3$) and transition ($B3$, $B4$) is a Higgs
transition, described by spinon $z^s_\alpha$ or $z^c_\alpha$
coupled with SU(2) CS theory in Eq.~\ref{su2cs}. The universality
class of these transitions is not understood yet.

\section{situation with $\mathrm{SU(2)_{charge}}$ broken to $\mathrm{U(1)_{charge}} \otimes Z_2$}

When the $\mathrm{SU(2)_{charge}}$ symmetry is broken down to
$\mathrm{U(1)_{charge}} \otimes Z_2$ symmetry which corresponds to
charge conservation and particle-hole transformation, the
degeneracy between $Q_{3b}$ and $Q_{1b}$, $Q_{2b}$ is lifted. For
instance, if an extra repulsive next nearest neighbor density
interaction (linear with $\mathcal{S}_{zz}$ in
Eq.~\ref{chargenematic}) is turned on \cite{raghu2008}, the system
favors to
develop $Q_{3b}$ %over $Q_{1b}$, $Q_{2b}$
$i.e.$ the system only has QSH order with GSM $S^2$. Then
according to Ref.~\cite{senthil2007}, the Skyrmion of the QSH
vector carries charge-$2e$, and Skyrmion condensate is a $s-$wave
superconductor.

If the system favors to have T-SC $Q_{1b}$, $Q_{2b}$ rather than
$Q_{3b}$, then depending on the microscopic parameters the T-SC
can have orders with either $Q_{1b}
\parallel Q_{2b}$ (type $A$) or $Q_{1b} \perp Q_{2b}$ (type $B$).
The type $A$ phase has fully gapped fermion spectrum, %it breaks
%the $[\mathrm{SU(2)_{spin} \otimes U(1)_{charge}}]/Z_2$ symmetry
%down to $\mathrm{U(1)_{spin}}$ symmetry,
with GSM$\sim [S^2 \otimes S^1]/Z_2$. Here $S^2$ corresponds to
the spin direction of the triplet Cooper pair, while $S^1$
corresponds to the pairing phase angle. Again, there are spin and
charge topological defects. For instance, the charge vortex
(defect of the $S^1$ part of the GSM) carries spin-1/2 quantum
number (quantum number of the residual $\mathrm{U(1)_{spin}}$
symmetry). The proliferation of the charge vortex leads to the
phase $A2$ in phase diagram Fig.~\ref{pd}, and the transition is a
CP(1) theory with easy plane anisotropy on charge CP(1) field
$z^c_\alpha$ introduced in Eq.~\ref{cp1}, which is consistent with
the conclusion in Ref.~\cite{ran2008}.

The proliferation of the spin-Skyrmion restores the
$\mathrm{SU(2)_{spin}}$ symmetry, but the $\mathrm{U(1)_{charge}}$
symmetry is still broken. However, since the spin and charge
sectors can change sign simultaneously without modifying the
ground state, after the proliferation of the spin-Skyrmion there
is still a $Z_2$ gauge symmetry for the charge manifold. Therefore
the GSM of this phase is $S^1/Z_2$, which corresponds to a
charge-$4e$ superconductor, instead of a charge-$2e$
superconductor. A similar scenario was discussed for the polar
state of the ultra-cold spin-1 spinor condensate
\cite{xumoore2006}, which also has the GSM $[S^2 \otimes
S^1]/Z_2$.

Unlike type $A$ order, the type $B$ phase with $Q_{1b} \perp
Q_{2b}$ does not have a fully gapped fermion spectrum. For
instance, with $\langle Q_{11} \rangle = \langle Q_{22} \rangle
\neq 0$, only spin-up is paired and gapped out, while spin-down is
not gapped. Since the fermion spectrum is gapless, the quantum
number of defects is no longer topologically stable.

\section{Summary and discussion}

In this work we have classified the QSH and T-SC states on the
honeycomb lattice using the SO(4) symmetry for a large class of
extended Hubbard models. By analyzing the quantum numbers of
topological defects, we obtained two different phase diagrams,
which were also confirmed by the Majorana liquid formalism. The
results of this paper can be straightforwardly generalized to
other bipartite lattice. Our formalism also predicts a phase with
both spin nematic and N\'{e}el order, sandwiched between a fully
gapped $Z_2 \otimes Z_2$ liquid phase and the ordinary N\'{e}el
order, which can be checked in future using the similar method as
Ref.~\cite{meng}.

The results we obtained in this work explicitly demonstrates the
spin-charge duality of the Hubbard model. For instance, in phase
diagram Fig.~\ref{pd}, spin and charge view each other as
topological defects. A similar spin-charge duality was applied to
the cuprates high temperature superconductor
\cite{weng,kouqiweng}, and a global phase diagram with both spin
and charge excitations was studied recently in
Ref.~\cite{yepeng2010}. We also note that other authors
\cite{ran2008,ran2010} have also studied the duality between spin
and charge with the presence of QSH order parameters, for instance
an easy plane version of spin-charge duality was identified as the
self-duality of the easy plane noncompact CP(1) theory, in a model
with inplane spin anisotropy. This duality led to a direct
transition between inplane N\'{e}el order and $d-$wave
superconductor. In our current work we showed that the generic
symmetry of the Hubbard model and the condensate of matrix order
parameter $Q$ in Eq.~\ref{Q} give us a complete and explicit
duality between spin and charge in interacting electrons.

%showed that this spin-charge duality becomes explicit and complete
%with the condensate of matrix order parameter $Q$ in Eq.~\ref{Q},
%in a large class of extended Hubbard models describing interacting
%electrons.

In both Fig.~\ref{pd} and Fig.~\ref{pd2} there is a multi-critical
point with $r_s = r_c = 0$. The multi-critical point in
Fig.~\ref{pd} was analyzed in Ref.~\cite{xs}, and for large enough
spinon number this multi-critical point is stable. Also, it has
been proposed that a similar multi-critical point is responsible
for the spin liquid behavior in material
$\kappa-\mathrm{(ET)_2Cu_2(CN)_3}$ on the triangular lattice
\cite{qixu}. The multi-critical point in Fig.~\ref{pd2} is more
complicated, we will leave this multi-critical point to future
study.

%\section{Appendix: $Z_2 \otimes Z_2$ liquid phase of $A3$}
%\label{appendix}

The author appreciates the very helpful discussions with Leon
Balents and Andreas Ludwig.

\bibliography{honey}

\end{document}